\overfullrule=0pt

\input harvmac

%
%

{\obeylines\gdef\startdisplay#1
  {\catcode`\^^M=5$$#1\halign\bgroup\indent##\hfil&&\qquad##\hfil\cr}}
\outer\def\enddisplay{\crcr\egroup$$}

\chardef\other=12
\def\ttverbatim{\begingroup \catcode`\\=\other \catcode`\{=\other
  \catcode`\}=\other \catcode`\$=\other \catcode`\&=\other
  \catcode`\#=\other \catcode`\%=\other \catcode`\~=\other
  \catcode`\_=\other \catcode`\^=\other
  \obeyspaces \obeylines \tt}
{\obeyspaces\gdef {\ }}  

\outer\def\begintt{$$\let\par=\endgraf \ttverbatim \parskip=0pt
  \catcode`\|=0 \rightskip=-5pc \ttfinish}
{\catcode`\|=0 |catcode`|\=\other 
  |obeylines 
  |gdef|ttfinish#1^^M#2\endtt{#1|vbox{#2}|endgroup$$}}

\catcode`\|=\active
{\obeylines\gdef|{\ttverbatim\spaceskip=.5em plus.25em minus.15em\let^^M=\ \let|=\endgroup}}%

\def\a{{\alpha}}

\def\l{{\lambda}}

\def\b{{\beta}}

\def\g{{\gamma}}

\def\d{{\delta}}
\def\e{{\epsilon}}

\def\half{{1\over 2}}
\def\p{{\partial}}

\def\t{{\theta}}

\def\({\left(}
\def\){\right)}
\def\cF{{\cal F}}

\Title{\vbox{\hbox{AEI 2010-127 }}}
{\vbox{
	\centerline{\bf PSS: A FORM Program to Evaluate Pure Spinor Superspace Expressions}
 }
}

\bigskip\centerline{Carlos R. Mafra\foot{email: crmafra@aei.mpg.de}}

\bigskip
\centerline{\it Max-Planck-Institut f\"ur
Gravitationsphysik} 
\smallskip
\centerline{\it Albert-Einstein-Institut, 14476 Golm, Germany}

\vskip .3in 

A {\tt FORM} program which is used to efficiently expand in components pure spinor 
superfield expressions of kinematic factors is presented and
comments on how it works are made. It is highly customizable using the standard features of {\tt FORM}
and can be used to help obtaining superstring effective actions from the scattering amplitudes
computed with the pure spinor formalism.

\vskip .3in

\Date {July 2010}

\lref\psf{
	N.~Berkovits,
	``Super-Poincare covariant quantization of the superstring,''
	JHEP {\bf 0004}, 018 (2000)
	[arXiv:hep-th/0001035].
}

\lref\multiloop{
  	N.~Berkovits,
  	``Multiloop amplitudes and vanishing theorems using the pure spinor
 	 formalism for the superstring,''
  	JHEP {\bf 0409}, 047 (2004)
  	[arXiv:hep-th/0406055].
} 
\lref\twoloop{
  	N.~Berkovits,
  	``Super-Poincare covariant two-loop superstring amplitudes,''
  	JHEP {\bf 0601}, 005 (2006)
  	[arXiv:hep-th/0503197].
}
\lref\brennotree{
	N.~Berkovits and B.~C.~Vallilo,
	``Consistency of super-Poincare covariant superstring tree amplitudes,''
	JHEP {\bf 0007}, 015 (2000)
	[arXiv:hep-th/0004171].
}

\lref\explsuper{
  N.~Berkovits,
  ``Explaining pure spinor superspace,''
  [arXiv:hep-th/0612021].
} 
\lref\PSequivII{
    N.~Berkovits and C.~R.~Mafra,
        ``Equivalence of two-loop superstring amplitudes in the pure spinor and RNS
        formalisms,''
        Phys.\ Rev.\ Lett.\  {\bf 96}, 011602 (2006)
        [arXiv:hep-th/0509234].
} 

\lref\oneloop{
  C.~R.~Mafra,
  ``Four-point one-loop amplitude computation in the pure spinor formalism,''
  JHEP {\bf 0601}, 075 (2006)
  [arXiv:hep-th/0512052].
} 

\lref\stahn{
  C.~Stahn,
  ``Fermionic superstring loop amplitudes in the pure spinor formalism,''
  JHEP {\bf 0705}, 034 (2007)
  [arXiv:0704.0015 [hep-th]].
}

\lref\mafraids{
  C.~R.~Mafra,
  ``Pure Spinor Superspace Identities for Massless Four-point Kinematic
  Factors,''
  JHEP {\bf 0804}, 093 (2008)
  [arXiv:0801.0580 [hep-th]].
}

\lref\NMPS{
    N.~Berkovits,
        ``Pure spinor formalism as an N = 2 topological string,''
    JHEP {\bf 0510}, 089 (2005)
    [arXiv:hep-th/0509120].
} 
\lref\anom{
  N.~Berkovits and C.~R.~Mafra,
  ``Some superstring amplitude computations with the non-minimal pure spinor
  formalism,''
  JHEP {\bf 0611}, 079 (2006)
  [arXiv:hep-th/0607187].
}

\lref\twolooptwo{
    N.~Berkovits and C.~R.~Mafra,
        ``Equivalence of two-loop superstring amplitudes in the pure spinor and RNS
        formalisms,''
        Phys.\ Rev.\ Lett.\  {\bf 96}, 011602 (2006)
        [arXiv:hep-th/0509234].
}

\lref\nekr{
  N.~Berkovits and N.~Nekrasov,
  ``Multiloop superstring amplitudes from non-minimal pure spinor formalism,''
  JHEP {\bf 0612}, 029 (2006)
  [arXiv:hep-th/0609012].
}
\lref\Rpuri{
    G.~Policastro and D.~Tsimpis,
        ``R**4, purified,''
        [arXiv:hep-th/0603165].
}
\lref\fivept{
  C.~R.~Mafra and C.~Stahn,
  ``The One-loop Open Superstring Massless Five-point Amplitude with the
  Non-Minimal Pure Spinor Formalism,''
  JHEP {\bf 0903}, 126 (2009)
  [arXiv:0902.1539 [hep-th]].
}
\lref\mafrabcj{
  C.~R.~Mafra,
  ``Simplifying the Tree-level Superstring Massless Five-point Amplitude,''
  JHEP {\bf 1001}, 007 (2010)
  [arXiv:0909.5206 [hep-th]].
}
\lref\WittenNT{
  E.~Witten,
  ``Twistor - Like Transform In Ten-Dimensions,''
  Nucl.\ Phys.\  B {\bf 266}, 245 (1986).
}
\lref\HarnadBC{
  J.~P.~Harnad and S.~Shnider,
  ``Constraints And Field Equations For Ten-Dimensional Superyang-Mills
  Theory,''
  Commun.\ Math.\ Phys.\  {\bf 106}, 183 (1986).
}

\lref\psspace{
  N.~Berkovits,
  ``Explaining pure spinor superspace,''
  [arXiv:hep-th/0612021].
}

\lref\vanh{
  L.~Anguelova, P.~A.~Grassi and P.~Vanhove,
  ``Covariant one-loop amplitudes in D = 11,''
  Nucl.\ Phys.\  B {\bf 702}, 269 (2004)
  [arXiv:hep-th/0408171].
}
\lref\gammaMat{
  U.~Gran,
  ``GAMMA: A Mathematica package for performing Gamma-matrix algebra and  Fierz
  transformations in arbitrary dimensions,''
  arXiv:hep-th/0105086.
}
\lref\FORM{
  J.~A.~M.~Vermaseren,
  ``New features of FORM,''
  [arXiv:math-ph/0010025]
 \semi
  M.~Tentyukov and J.~A.~M.~Vermaseren,
  ``The multithreaded version of FORM,''
  [arXiv:hep-ph/0702279].
}
\lref\mafraFT{
  C.~R.~Mafra,
  ``Towards Field Theory Amplitudes From the Cohomology of Pure Spinor
  Superspace,''
  arXiv:1007.3639 [hep-th].
}
\lref\Sdual{
  E.~D'Hoker, M.~Gutperle and D.~H.~Phong,
  ``Two-loop superstrings and S-duality,''
  Nucl.\ Phys.\  B {\bf 722}, 81 (2005)
  [arXiv:hep-th/0503180].
}
\lref\twocoeff{
  H.~Gomez and C.~R.~Mafra,
  ``The Overall Coefficient of the Two-loop Superstring Amplitude Using Pure
  Spinors,''
  JHEP {\bf 1005}, 017 (2010)
  [arXiv:1003.0678 [hep-th]].
}
\lref\humberto{
  H.~Gomez,
  ``One-loop Superstring Amplitude From Integrals on Pure Spinors Space,''
  JHEP {\bf 0912}, 034 (2009)
  [arXiv:0910.3405 [hep-th]].
}
\lref\vanhoveBCJ{
  N.~E.~J.~Bjerrum-Bohr, P.~H.~Damgaard, T.~Sondergaard and P.~Vanhove,
  ``Monodromy and Jacobi-like Relations for Color-Ordered Amplitudes,''
  arXiv:1003.2403 
}
\lref\Medinas{
  R.~Medina, F.~T.~Brandt and F.~R.~Machado,
  ``The open superstring 5-point amplitude revisited,''
  JHEP {\bf 0207}, 071 (2002)
  [arXiv:hep-th/0208121].
\semi
  L.~A.~Barreiro and R.~Medina,
  ``5-field terms in the open superstring effective action,''
  JHEP {\bf 0503}, 055 (2005)
  [arXiv:hep-th/0503182].
}
\lref\thetaSYM{
  	J.~P.~Harnad and S.~Shnider,
	``Constraints And Field Equations For Ten-Dimensional Superyang-Mills
  	Theory,''
  	Commun.\ Math.\ Phys.\  {\bf 106}, 183 (1986).
\semi
	P.~A.~Grassi and L.~Tamassia,
        ``Vertex operators for closed superstrings,''
        JHEP {\bf 0407}, 071 (2004)
        [arXiv:hep-th/0405072].
}
\lref\mukho{
  P.~Mukhopadhyay,
  ``On D-brane boundary state analysis in pure-spinor formalism,''
  JHEP {\bf 0603}, 066 (2006)
  [arXiv:hep-th/0505157].
}
\lref\bcj{
  Z.~Bern, J.~J.~M.~Carrasco and H.~Johansson,
  ``New Relations for Gauge-Theory Amplitudes,''
  Phys.\ Rev.\  D {\bf 78}, 085011 (2008)
  [arXiv:0805.3993 [hep-ph]].
}
\lref\bohrbcj{
  N.~E.~J.~Bjerrum-Bohr, P.~H.~Damgaard and P.~Vanhove,
  ``Minimal Basis for Gauge Theory Amplitudes,''
  Phys.\ Rev.\ Lett.\  {\bf 103}, 161602 (2009)
  [arXiv:0907.1425 [hep-th]].
}
\lref\stiebcj{
  S.~Stieberger,
  ``Open \& Closed vs. Pure Open String Disk Amplitudes,''
  arXiv:0907.2211 
}
\lref\stieMHV{
  S.~Stieberger and T.~R.~Taylor,
  ``Amplitude for N-gluon superstring scattering,''
  Phys.\ Rev.\ Lett.\  {\bf 97}, 211601 (2006)
  [arXiv:hep-th/0607184].
  \semi
  S.~Stieberger and T.~R.~Taylor,
  ``Multi-gluon scattering in open superstring theory,''
  Phys.\ Rev.\  D {\bf 74}, 126007 (2006)
  [arXiv:hep-th/0609175].
  \semi
  S.~Stieberger and T.~R.~Taylor,
  ``Supersymmetry Relations and MHV Amplitudes in Superstring Theory,''
  Nucl.\ Phys.\  B {\bf 793}, 83 (2008)
  [arXiv:0708.0574 [hep-th]].
  \semi
  S.~Stieberger and T.~R.~Taylor,
  ``Complete Six-Gluon Disk Amplitude in Superstring Theory,''
  Nucl.\ Phys.\  B {\bf 801}, 128 (2008)
  [arXiv:0711.4354 [hep-th]].
}
\lref\ggi{
  O.~A.~Bedoya and N.~Berkovits,
  ``GGI Lectures on the Pure Spinor Formalism of the Superstring,''
  arXiv:0910.2254 [hep-th].
}
\lref\ictp{
  N.~Berkovits,
  ``ICTP lectures on covariant quantization of the superstring,''
  arXiv:hep-th/0209059.
}
\lref\tese{
  C.~R.~Mafra,
  ``Superstring Scattering Amplitudes with the Pure Spinor Formalism,''
  arXiv:0902.1552 [hep-th].
}

\newsec{Introduction}

Since the discovery of the pure spinor formalism \psf\
the computation of manifestly supersymmetric superstring scattering 
amplitudes became possible\foot{For reviews see \refs{\ictp,\ggi,\tese}.}.
At first the results were limited to tree-level,
where it was shown that amplitudes with an arbitrary number
of bosonic and up to four fermionic states were equivalent
to the standard results from the Ramond-Neveu-Schwarz (RNS) and Green-Schwarz (GS)
formulations \brennotree. Explicit four-point tree computations were first performed
in \Rpuri, while in \mafraids\ those results were streamlined in a superspace derivation which
also made manifest its relation with one- and two-loop amplitudes.
The five-point amplitude was computed in \mafrabcj, providing a compact superspace
representation which contrasts with the bosonic-only result from \Medinas. In addition, an OPE
identity related to the Bern-Carrasco-Johansson kinematic relations \bcj\ was uncovered\foot{See also
\refs{\bohrbcj,\stiebcj} for string theory monodromy explanations of the BCJ relations.}, which led to
further developments discussed  in \vanhoveBCJ. Furthermore, it was shown in \mafraFT\ that there is
a BRST-equivalent superspace expression for the field theory limit of the superstring amplitude
from \mafrabcj\ which provides hints of a direct mapping between Feynman diagrams with cubic vertices
and pure spinor superspace expressions.

After being extended in \refs{\multiloop,\NMPS},
higher-loop amplitude computations using the pure spinor formalism also became a reality.
At one-loop, the massless four-point \refs{\multiloop,\oneloop}, five-point \fivept\ and the
gauge variation of the six-point amplitude \anom\ were obtained. 
At two-loops, the massless four-point amplitude was computed in \refs{\twoloop,\twolooptwo}. 
Using formul{\ae} for integration over pure spinor space, the overall coefficients of the one-loop \humberto\
and two-loop \twocoeff\ were also computed and shown to agree with S-duality conjecture 
expectations \Sdual.

Besides streamlining amplitude computations
avoiding supermoduli spaces and sums over spins structures altogether,
the pure spinor formalism naturally gives rise to manifestly supersymmetric
kinematic factors in {\it pure spinor superspace}.

Pure spinor superspace expressions are correlation functions 
written in terms of ten-dimensional super-Yang-Mills superfields \WittenNT\ 
and three pure spinors $\l^\a$ 
normalized as\foot{The precise
overall coefficients of \refs{\humberto,\twocoeff} will not be needed here.}
\eqn\norm{
\langle (\l\g^m \t)(\l\g^n \t)(\l\g^p \t)(\t\g_{mnp} \t)\rangle = 1.
}
The simplest example of a pure spinor superspace expression 
is provided by the massless three-point scattering amplitude \psf
\eqn\three{
K^{(0)}_3 = \langle (\l A^1)(\l A^2)(\l A^3)\rangle.
}
The four-point kinematic factors are given by
\eqn\tree{
K^{(0)}_4  = \langle (\l A^1)(\l\g^m W^2)(\l \g^n W^3){\cal F}^4_{mn}\rangle,
}
\eqn\Koneloop{
K^{(1)}_4 = \langle (\l A^1)(\l\g^m W^2)(\l \g^n W^3){\cal F}^4_{mn}\rangle,
}
\eqn\twoloop{
K^{(2)}_4 = 
\langle(\l\g^{mnpqr}\l)(\l\g^s W^4){\cal F}^1_{mn}{\cal F}^2_{pq}{\cal F}^3_{rs}\rangle
}
for the tree-level \mafraids, one- \multiloop\ and two-loop \twoloop\ amplitudes, respectively.
Another example is provided by the one-loop
five-point kinematic factor of \fivept, whose
expression for the $(12)$ and $(25)$ ``channels'' read
$$
\eqalign{
  L_{12} &= \langle
  \Big[(\l A^1)(k^1\cdot A^2) + A^1_p (\l\g^p W^2) \Big]
  (\l\g^m W^5)(\l\g^n W^3){\cal F}^4_{mn} \rangle \cr
  K_{25} &=  \langle
  (\l A^1) \Big[
  (\l\g^m W^2)(k^2\cdot A^5) - {1\over 4} (\l\g^m\g^{rs} W^5) {\cal F}^2_{rs} \Big]
  (\l\g^n W^3){\cal F}^4_{mn} \Bigr\rangle - (2\leftrightarrow 5).
}
$$

The above pure spinor superspace representations provide compact information about the
amplitudes, but it may be convenient to evaluate those expressions in
terms of familiar component expansions. These component expansions are
written in terms of polarization vectors $e_m^I$ and spinors $\chi^\a_I$ with
momenta $k_m^I$, where $I=1,{\ldots}, N$ are the particle labels and $m=0,{\ldots} ,9\; \a = 1,{\ldots} ,16$
are the Lorentz and Weyl indices of ten dimensional Minkowski space. 

The general method to evaluate these expressions in components was explained
in the appendix of \anom. One uses the
$\t$-expansions \refs{\thetaSYM,\Rpuri} of the SYM superfields
$$
A_{\a}(x,\t)={1\over 2}a_m(\g^m\t)_\a -{1\over 3}(\xi\g_m\t)(\g^m\t)_\a
-{1\over 32}F_{mn}(\g_p\t)_\a (\t\g^{mnp}\t) + \ldots
$$
$$
A_{m}(x,\t) = a_m - (\xi\g_m\t) - {1\over 8}(\t\g_m\g^{pq}\t)F_{pq}
         + {1\over 12}(\t\g_m\g^{pq}\t)(\p_p\xi\g_q\t) + \ldots
$$
$$
W^{\a}(x,\t) = \xi^{\a} - {1\over 4}(\g^{mn}\t)^{\a} F_{mn}
           + {1\over 4}(\g^{mn}\t)^{\a}(\p_m\xi\g_n\t)
	   + {1\over 48}(\g^{mn}\t)^{\a}(\t\g_n\g^{pq}\t)\p_m F_{pq} 
	   + \ldots
$$
\eqn\expansions{
\cF_{mn}(x,\t) = F_{mn} - 2(\p_{[m}\xi\g_{n]}\t) + {1\over
4}(\t\g_{[m}\g^{pq}\t)\p_{n]}F_{pq} + {\ldots}
}
where $a_m(x) = e_m {\rm e}^{ikx}$ and $\xi^\a(x)=\chi^\a {\rm e}^{ikx}$. After that,
only terms containing five $\t$'s are kept. Using symmetry alone it is
possible to rewrite arbitrary combinations of $\langle \l^3 \t^5\rangle$ in terms
of Kronecker deltas and epsilon tensors \refs{\twolooptwo,\stahn}. For example
\eqn\example{
\langle (\l\g^m\t)(\l\g^n\t)(\l\g^p \t)(\t\g_{qrs}\t)\rangle = {1\over 120}\d^{mnp}_{qrs}.
}
Substituting the various correlators by their corresponding tensors as above, the
component expansion of pure spinor superspace expressions is obtained.

Several different computer-aided
procedures were used along the past years for the above steps in superspace derivations of
scattering amplitudes.
At first, those pure spinor correlators of \anom\ were obtained
with the help\foot{A small C program was also
written to deal with anti-symmetrization of huge tensors during consistency checks.} 
of the {\tt GAMMA} package \gammaMat.
With some effort, the superspace expressions were expanded in $\t$'s by hand,
the corresponding correlators identified from a catalog composed of entries like \example\ 
and the resulting tensors were typed in Mathematica or {\tt FORM} \FORM. When the scattering involved fermionic states,
various Fierz identities were usually necessary at intermediate stages. After all these steps,
the output consisted of several terms composed of Kronecker deltas and Levi-Civita epsilon tensors contracted
with momenta and polarizations. Those terms constituted the final answer.

However the semi-automated method described above does not scale very well 
for higher-point amplitudes or when there are many expressions to evaluate in sequence.
This fact provided the motivation to write the program presented here.
It was developed mainly to help the author's own workflow during computations, 
and therefore it reflects his priorities (and it is a continuous work in progress).
It is called {\tt PSS} and it is written in the interesting language of {\tt FORM} \FORM.

\newsec{How PSS works}

The goal is to be able to obtain component expansions 
in a fully automated process --
all that is required is the pure spinor superspace expression to be expanded and whether
the external states are bosonic or fermionic. The rest must be done by the computer.

For example, the
superspace expression for \oneloop\ is typed in {\tt PSS} as
\begintt
 Local [4-pts_one-loop] = la*A1*la*ga(m)*W2*la*ga(n)*W3*cF4(m,n);
\endtt 
Choosing all fields to be bosonic results in the following {\it ipsis litteris} output,
\begintt
   [4-pts_one-loop] =
       - 1/5760*k1.e2*k1.e3*e1.e4*t
       + 1/5760*k1.e2*k1.e4*e1.e3*t
       + 1/5760*k1.e2*k2.e3*e1.e4*u
       + 1/5760*k1.e2*k2.e4*e1.e3*t
       + 1/5760*k1.e2*k3.e1*e3.e4*t
       + 1/5760*k1.e3*k2.e1*e2.e4*t
       - 1/5760*k1.e3*k2.e4*e1.e2*t
       - 1/5760*k1.e3*k3.e2*e1.e4*u
       - 1/5760*k1.e3*k3.e2*e1.e4*t
       + 1/5760*k1.e4*k2.e1*e2.e3*u
       - 1/5760*k1.e4*k2.e3*e1.e2*u
       + 1/5760*k1.e4*k3.e2*e1.e3*u
       + 1/5760*k1.e4*k3.e2*e1.e3*t
       - 1/5760*k2.e1*k2.e3*e2.e4*u
       + 1/5760*k2.e1*k2.e4*e2.e3*u
       + 1/5760*k2.e1*k3.e2*e3.e4*u
       - 1/5760*k2.e3*k3.e1*e2.e4*u
       - 1/5760*k2.e3*k3.e1*e2.e4*t
       + 1/5760*k2.e4*k3.e1*e2.e3*u
       + 1/5760*k2.e4*k3.e1*e2.e3*t       
       + 1/5760*k3.e1*k3.e2*e3.e4*u
       + 1/5760*k3.e1*k3.e2*e3.e4*t
       + 1/11520*e1.e2*e3.e4*t*u
       - 1/11520*e1.e3*e2.e4*t*u
       - 1/11520*e1.e3*e2.e4*t^2
       - 1/11520*e1.e4*e2.e3*u^2
       - 1/11520*e1.e4*e2.e3*t*u
      ;

Momentum conservation: k4 eliminated
Gauge invariance: not tested 

  0.05 sec + 0.08 sec: 0.13 sec out of 0.15 sec
\endtt
which is the result obtained in \oneloop. One should notice in the final 
statistics displayed by {\tt FORM} how quickly the answer
is obtained.

The program is composed of one main FORM script called {\tt pss.frm}
and four header files: {\tt pss\_header.h}, {\tt kin\_factor.h}, {\tt pss.h} and
{\tt ps\_tensors.h}. They contain
the definitions of indices, vectors, tensors, superfields etc ({\tt pss\_header.h}),
the pure spinor superspace expressions to be evaluated ({\tt kin\_factor.h}) and
the procedures which actually do the computations ({\tt pss.h}). The database of 
pure spinor correlators is contained in {\tt ps\_tensors.h}.
There is also a small {\tt sed} script  (FORM2tex.sed) to help translating the result into $\TeX$.

To use the program one has to write down the kinematic factor in {\tt kin\_factor.h}.
In the beginning of the main file {\tt pss.frm}, the number of points must be defined 
(e.g. {\tt \#define Npts ``4''}) and whether the external states are bosonic or fermionic 
({\tt \#define field1 ``0''} and so forth). After that one executes {\tt pss.frm} using
either form or tform (for multi-processor computers) and a result like the one written above is obtained.
Optionally one can select which momentum to be eliminated by setting the dollar variable {\tt \$kn}
and test for gauge invariance by uncommenting the line containing {\tt id e1 = k1; \#gauge = e1;} close to the end
of the file. Several other things can be done, depending on the problem at hand and how one chooses
to manipulate it. There are also a few debug options (-d {psonly, sfexpand, nofierz}) which
help in case something goes wrong and one has to check where. They will be explained below.

\subsec{User input and notation}

The superspace expression is written in terms of the super-Yang-Mills superfields
$A^I_\a(\t,x)$, $A^I_m(\t,x)$, $W^\a_I(\t,x)$ and ${\cal F}^I_{mn}(\t,x)$ \WittenNT.
Their definitions are contained in the file {\tt pss\_header.h} and they correspond to,
\begintt A1, B1, W1, cF1(m,n)
\endtt
The pure spinor
$\l^\a$  and $\t^\a$ are denoted by {\tt la} and {\tt th}. Note that PSS does not
know about spinor indices, but that does not cause problems as long as one writes down 
correct superspace expressions. However, that also means one has to take care
whether a fermionic superfield is contracted from the left or right, because the
$\t$-expansions to be used differ in this case\foot{This is an artifact
of how PSS was designed and this distinction is meaningless in real life.}. For this
situation one has to use the ``left'' version of the superfields, AL1 and WL1 etc.
For example, the factor $(W^1\g_{mnp} W^2)$ must be written as 
\begintt WL1*ga(m,n,p)*W2
\endtt
The $16\times 16$ gamma matrices $\g^{m},\g^{mn}, \g^{mnp},{\ldots}$ 
are denoted by 
\begintt ga(m), ga(m,n), ga(m,n,p), ...
\endtt
Another convention to be followed is to write the pure spinor $\l^\a$ to the left of
fermionic bilinears,
so $(\l\g^m W^2)$ or $(W^2 \g^m \l)$ must be written as \begintt la*ga(m)*W2
\endtt
and not \begintt WL2*ga(m)*la
\endtt
although PSS can be easily modified to accept the latter version.
Of course, there is no problem to write the factor $(\l\g^{mnpqr}\l)$ and the
procedure to identify correlators is aware of it. The generalized
Kronecker delta is defined by\foot{Unfortunately {\tt FORM}
has no notion of a generalized Kronecker delta, so it had to be defined
by PSS. Note that the usual Kronecker delta
is defined by {\tt FORM} as $\d^m_n =$ {\rm d\_}(m,n).}
$$
N! \d^{m_1{\ldots} m_N}_{n_1{\ldots} n_N} = {\rm da}(m_1,{\ldots} ,m_N,n_1,{\ldots} ,n_N)
$$

\subsec{The computation}

It is important to understand how PSS actually obtains the component expansions,
so that modifications can be easily done. All the action happens inside {\tt pss.frm},
where it calls the procedures from {\tt pss.h}. Let us now follow some of the
steps\foot{
One should also read the source code, as some details will be skipped.}.

When PSS is executed it loads the headers and the kinematic factors. The kinematic
factors are local variables which will be manipulated by the {\tt FORM} program.
The first part of the manipulations transform the superspace input into an
expression suitable for identifying the required pure spinor
correlators in terms of tensors to arrive at the final answer. 
The (slightly simplified) sequence of procedures is the
following:
\begintt
  #call chooseMomentum()
  #call superfieldExpand()
  #call keep5thetas()
  #call gammaExpand()
  #call deltaExpand()
  #call onShell()
  #call orderFermions()
  #call PSordering()
  #call Fierz()
  #call earlySimplify()
  #call identifyCorrelators()
  #call deltaExpand()
  #call dualizeGammas()
  #call fieldStrength()
  #call onShell()
  #call gammaExpand()
  #call orderFermions()
  #call diracEquation()
  #call momentumConservation()
\endtt
These steps are almost self-explanatory, and they correspond to what
one would actually do in a computation with pen and paper.

The procedure {\tt chooseMomentum()} is called to choose which momentum 
to eliminate. Although the result does not depend on the choice, there are
differences in the number of intermediate terms and
computing time.
For example, when there are fermionic
fields it is convenient to eliminate one momentum whose label is not one
of the labels of the fermionic particles. The reason is the increased
chance of applying the Dirac equation on fermion bilinears to reduce 
the rank of the gamma matrix. For example,
$(\chi^1\g^{mnp}\chi^2)k^1_m = 2(\chi^1 \g^{[n} \chi^2)k_1^{p]}$.

Then the procedure {\tt superfieldExpand()} is called, which expands the
SYM superfields in their $\t$ components using \expansions\ and selects the terms
according to whether the particles are bosonic or fermionic.
The assignment of particles to each label of the superfields 
is done at the
beginning of {\tt pss.frm} with
\begintt
#define field1 ``0''
#define field2 ``0''
...
\endtt
where $0$ ($1$) means bosonic (fermionic).
Terms which do not contain five $\t$s are then discarded with
{\tt keep5thetas()}.
At this point, one of the
terms in \Koneloop\ looks like
\begintt
- 1/256*la*ga(e1)*th*la*ga(k4)*ga(m2,n2)*th*la*ga(n)*ga(m3,n3)*th*
        th*ga(n,m4,n4)*th*F2(m2,n2)*F3(m3,n3)*F4(m4,n4)
\endtt      
which is the {\tt FORM} output for
\eqn\iterme{
-{1\over 256}\langle (\l\g^{m}\t)(\l\g^{p}\g^{m_2n_2}\t)(\l\g^n\g^{m_3n_3}\t)
(\t\g_{nm_4n_4}\t)\rangle k^4_p e^1_m F^2_{m_2n_2}F^3_{m_3n_3}F^4_{m_4n_4}
}
in the schoonschip notation.
The gamma matrices in \iterme\ are expanded using {\tt gammaExpand()}, 
$$
\g^n\g^{m_3n_3} = \g^{n m_3n_3} + \eta^{m_3n}\g^{n_3} -  \eta^{n_3n}\g^{m_3}.
$$
More complicated expansions may introduce generalized Kronecker deltas, which
are then expanded with {\tt deltaExpand()}, e.g.,
$\d^{mn}_{pq} = \half(\d^m_p \d^n_q - \d^m_q\d^n_p)$.
The procedure {\tt onShell()} kills any term which may have been generated 
at this point containing $(k^i\cdot k^i)$
or $(k^i\cdot e^i)$.

When there are fermionic external states, the procedure {\tt orderFermions()}
rewrites the fermionic bilinears
$(\chi^i\g^{m_1{\ldots}m_n }\chi^j)$ such that
$i < j$. This procedure keeps track of overall minus signs which 
may be needed due to the Grassmanian nature of $\chi$'s and the
symmetry properties of the gamma matrices.
For example $(\chi^3\g^m \chi^1) \rightarrow - (\chi^1 \g^m \chi^3)$
or $(\chi^3\g^{mnp} \chi^1) \rightarrow + (\chi^1 \g^{mnp} \chi^3)$.

The procedure {\tt PSordering()} follows a set of conventions
on the ordering of fermionic bilinears to minimize the number of pattern
matching when trying to identify the pure spinor correlators needed for
one particular computation. For example, if the expression $(\l \g^m \t)(\t \g_{qrs}\t)(\l\g^n\t)(\l\g^p\t)$
is encountered in the middle of the computations, it is first rewritten by {\tt PSordering()} 
as $(\l \g^m \t)(\l\g^n\t)(\l\g^p\t)(\t \g_{qrs}\t)$. Later on 
there will be only one pattern to match in order to identify the correlator 
(in {\tt identifyCorrelators()})
and replace it with ${1\over 720}${\rm da}(m,n,p,q,r,s). For the same reason, the ordering ``inside''
gamma matrix bilinears is such that the fields appear as $(\l\g^{m{\ldots} }\t)$,
$(\l\g^{m{\ldots} }\chi)$ and $(\chi\g^{m{\ldots} }\t)$, and not for example as
$(\t\g^{m{\ldots} }\l)$ etc.

When the computation involves fermionic particles, after processing the expressions with
{\tt PSordering()}
there may be factors such as $(\chi^1 \g_{m}\t)(\chi^2 \g^n \t)$, which will be rewritten
by the procedure {\tt Fierz()} as 
$$
- {1\over 96}(\chi^1\g^m\g_{qrs}\g^n \chi^2)(\t \g^{qrs}\t)
$$
again in order to minimize the number of pattern matching to identify correlators. This procedure
takes care of various bilinears combinations, and the pair of fermions chosen to be expanded
depend on the particular combination being considered, for which the user should read the
source code for more information. The general expansions are done with the formula
\eqn\Fierz{
\l^\a \chi^\b = {1\over 16}(\l\g^m \chi)\g_m^{\a\b} + {1\over 96}(\l\g^{mnp}\chi)\g_{mnp}^{\a\b}
+ {1\over 3840}(\l\g^{mnpqr}\chi)\g_{mnpqr}^{\a\b}.
}
If the procedure {\tt Fierz()} takes effect, then there will be more gamma matrices to expand,
like in the example above $(\chi^1\g^m\g_{qrs}\g^n \chi^2)$. If that is the case, then
the extra calling of {\tt gammaExpand()} and {\tt deltaExpand()}, followed by {\tt PSordering()}
and {\tt orderFermions()} will let the expressions ready for being identified. 

Using the example of the one-loop kinematic factor, at this point one of the terms
being dealt with by {\tt PSS} is 
\begintt
- 1/512*la*ga(N1_?,N2_?,N3_?)*th*la*ga(N4_?,N5_?,N6_?)*th*
  la*ga(N7_?)*th*th*ga(N7_?,N8_?,N9_?)*th*F1(N8_?,N9_?)*
  F2(N1_?,N2_?)*F3(N4_?,N5_?)*F4(N3_?,N6_?)
\endtt
which is
\eqn\corre{
-{1\over 512}(\l\g^{n_1n_2n_3}\t)(\l\g^{n_4n_5n_6}\t)(\l\g^{n_7}\t)(\t\g^{n_7n_8n_9}\t)F^1_{n_8n_9}F^2_{n_1n_2}F^3_{n_4n_5}F^4_{n_3n_6}.
}
The next step is to call the procedure which identifies the pure spinor correlator from a catalog of 
known tensors. This is done with {\tt identifyCorrelators()}, after which the above term is given
by
\begintt
 - 1/512*F1(N1_?,N2_?)*F2(N3_?,N4_?)*F3(N5_?,N6_?)*F4(N7_?,N8_?)*
   ps331(N3_?,N4_?,N7_?,N5_?,N6_?,N8_?,N9_?,N9_?,N1_?,N2_?)
\endtt
That is, {\tt PSS} identifies the pure spinor correlator $(\l\g^{n_1n_2n_3}\t)(\l\g^{n_4n_5n_6}\t)(\l\g^{n_7}\t)(\t\g^{n_7n_8n_9}\t)$
with the tensor ${\rm ps331}(n_1,n_2,n_3,n_4,n_5,n_6,n_7,n_7,n_8,n_9)$. Looking at the appendix of \anom, this
tensor is expanded in terms of Kronecker deltas as
\eqn\tres{
\langle(\l\g^{mnp}\t)(\l\g^{qrs}\t)(\l\g_{t}\t)(\t\g_{ijk}\t)\rangle=
{1\over 8400}\e^{ijkmnpqrst}+
}
$$
+{1\over 140}\Big[ 
	 \d^{[m}_t\d^n_{[i}\eta^{p][q}\d^r_j\d^{s]}_{k]}
	-\d^{[q}_t\d^r_{[i}\eta^{s][m}\d^n_j\d^{p]}_{k]}
\Big]
-{1\over 280}\Big[ 
	 \eta_{t[i}\eta^{v[q} \d^r_j\eta^{s][m}\d^n_{k]}\d^{p]}_v 
	-\eta_{t[i}\eta^{v[m} \d^n_j\eta^{p][q}\d^r_{k]}\d^{s]}_v
\Big]
$$
and this is one of the correlators included in the catalog {\tt ps\_tensors.h}.
So the purpose of {\tt identifyCorrelators()} is to transform an input containing
correlators with $\langle \l^3\t^5\rangle$ into an expression written in terms
of tensors like {\rm ps331(m,n,p,q,r,s,t,u,v,x)}, which are later substituted by
Kronecker deltas and epsilon tensors as in \tres\ using the catalog.

The list of correlators identified by {\tt identifyCorrelators()} is
$$
\langle (\l \g^{t_1...t_5} \l)
        (\l \g^{m}\t)
        (\t \g^{r_1...r_3} \t)
        (\t \g^{s_1...s_3} \t) \rangle = 
     {\rm uind}(t_1,...,t_5,m,r_1,...,r_3,s_1,...,s_3)
$$
$$
     \langle (\l \g^{t_1...t_5} \l)
        (\l \g^{m_1...m_3} \t)
        (\t \g^{r_1...r_3} \t)
        (\t \g^{s_1...s_3} \t) \rangle = 
     {\rm tind}(t_1,...,t_5,m_1,...,m_3,r_1,...,r_3,s_1,...,s_3)
$$     
$$
     \langle (\l \g^{t_1...t_5} \l)
        (\l \g^{m_1...m_5} \t)
        (\t \g^{r_1...r_3} \t)
        (\t \g^{s_1...s_3} \t) \rangle = 
     {\rm cind}(t_1,...,t_5,m_1,...,m_5,r_1,...,r_3,s_1,...,s_3)
$$
$$
       \langle (\l \g^{m} \t)
          (\l \g^{n} \t)
          (\l \g^{p} \t)
          (\t \g^{abc} \t) \rangle = 
       {\rm ps111}(m,n,p,a,b,c)
$$
$$
       \langle (\l \g^{m} \t)
          (\l \g^{n} \t)
          (\l \g^{p} \t)
          (\t \g^{r_1...r_7} \t) \rangle = 
       {\rm ps111eps}(m,n,p,r_1,...,r_7)
$$
$$
       \langle (\l \g^{t_1...t_3} \t)
          (\l \g^{m} \t)
          (\l \g^{n} \t)
          (\t \g^{r_1...r_3} \t) \rangle = 
       {\rm ps311}(t_1,...,t_3,m,n,r_1,...,r_3)
$$
$$
       \langle (\l \g^{t_1...t_3} \t)
          (\l \g^{m} \t)
          (\l \g^{n} \t)
          (\t \g^{r_1...r7} \t) \rangle = 
       {\rm ps311eps}(t_1,...,t_3,m,n,r_1,...,r7)
$$
$$
       \langle (\l \g^{m_1...m_3} \t)
          (\l \g^{n_1...n_3} \t)
          (\l \g^{a} \t)
          (\t \g^{r_1...r_3} \t) \rangle = 
       {\rm ps331}(m_1,...,m_3,n_1,...,n_3,a,r_1,...,r_3)
$$
$$
       \langle (\l \g^{m_1...m_3} \t)
          (\l \g^{n_1...n_3} \t)
          (\l \g^{a} \t)
          (\t \g^{r_1...r_7} \t) \rangle = 
       {\rm ps331eps}(m_1,...,m_3,n_1,...,n_3,a,r_1,...,r_7)
$$
$$
       \langle (\l \g^{t_1...t_3} \t)
          (\l \g^{m_1...m_3} \t)
          (\l \g^{n_1...n_3} \t)
          (\t \g^{r_1...r_3} \t) \rangle = 
       {\rm ps333}(t_1,.,t_3,m_1,.,m_3,n_1,.,n_3,r_1,.,r_3)
$$
$$
       \langle (\l \g^{t_1...t_3} \t)
          (\l \g^{m_1...m_3} \t)
          (\l \g^{n_1...n_3} \t)
          (\t \g^{r_1...r_7} \t) \rangle = 
       {\rm ps333eps}(t_1,.,t_3,m_1,.,m_3,n_1,.,n_3,r_1,.,r_7)
$$
$$
       \langle (\l \g^{t_1...t_5} \t)
          (\l \g^{m} \t)
          (\l \g^{a} \t)
          (\t \g^{r_1...r_3} \t) \rangle = 
       {\rm ps511}(t_1,.,t_5,m,a,r_1,.,r_3)
$$
$$
       \langle (\l \g^{t_1...t_5} \t)
          (\l \g^{m} \t)
          (\l \g^{a} \t)
          (\t \g^{r_1...r_7} \t) \rangle = 
       {\rm ps511eps}(t_1,.,t_5,m,a,r_1,.,r_7)
$$
$$
       \langle (\l \g^{t_1...t_5} \t)
          (\l \g^{m_1...m_3} \t)
          (\l \g^{a} \t)
          (\t \g^{r_1...r_3} \t) \rangle = 
       {\rm ps531}(t_1,.,t_5,m_1,.,m_3,a,r_1,.,r_3)
$$
$$
       \langle (\l \g^{t_1...t_5} \t)
          (\l \g^{m_1...m_3} \t)
          (\l \g^{a} \t)
          (\t \g^{r_1...r_7} \t) \rangle = 
       {\rm ps531eps}(t_1,...,t_5,m_1,...,m_3,a,r_1,...,r_7)
$$
$$
       \langle (\l \g^{t_1...t_5} \t)
          (\l \g^{m_1...m_5} \t)
          (\l \g^{a} \t)
          (\t \g^{r_1...r_3} \t) \rangle = 
       {\rm ps551}(t_1,...,t_5,m_1,...,m_5,a,r_1,...,r_3)
$$
$$
       \langle (\l \g^{t_1...t_5} \t)
          (\l \g^{m_1...m_5} \t)
          (\l \g^{a} \t)
          (\t \g^{r_1...r_7} \t) \rangle = 
       {\rm ps551eps}(t_1,.,t_5,m_1,.,m_5,a,r_1,.,r_7)
$$
$$
       \langle (\l \g^{t_1...t_5} \t)
          (\l \g^{m_1...m_5} \t)
          (\l \g^{n_1...n_3} \t)
          (\t \g^{r_1...r_3} \t) \rangle = 
       {\rm ps553}(t_1,.,t_5,m_1,.,m_5,n_1,.,n_3,r_1,.,r_3)
$$
$$
       \langle (\l \g^{t_1...t_5} \t)
          (\l \g^{m_1...m_3} \t)
          (\l \g^{n_1...n_3} \t)
          (\t \g^{r_1...r_3} \t) \rangle = 
       {\rm ps533}(t_1,.,t_5,m_1,.,m_3,n_1,.,n_3,r_1,.,r_3)
$$
It may happen that one particular computation requires a correlator not
in the list. When that happens {\tt PSS} automatically detects the missing
correlator and prints it before exiting. For example
\begintt
I do not identify a correlator for:
 + 1/983040*chi1*ga(N1_?,N2_?,N3_?,N4_?,N5_?,N6_?,N7_?)*chi2*
   la*ga(N1_?,N2_?,N3_?,N4_?,N5_?)*th*la*ga(k1,N6_?,N8_?)*th*
   la*ga(N9_?,N10_?,N11_?)*th*
   th*ga(N7_?,N8_?,N9_?,N12_?,N13_?,N14_?,N15_?)*th*
   F3(N10_?,N11_?)*F4(N12_?,N13_?)*F5(N14_?,N15_?)
Add it to the identifyCorrelators() procedure
\endtt
In this case the missing correlator can be obtained from {\rm ps533()} and its pattern matching
added
to the procedure {\tt identifyCorrelators()} and its tensor representation added to the file
{\tt ps\_tensors.h}.

After calling {\tt identifyCorrelators()} the tensors above are substituted by their
tensor representations. This is done by including the file {\tt ps\_tensors.h}, which contains
their expansions in terms of generalized Kronecker deltas and Levi-Civita epsilons.
The procedure {\tt deltaExpand()} expands the generalized Kronecker deltas in terms of
the antisymmetric combinations of the usual Kronecker delta. Gamma matrices with more than
five indices are manipulated with {\tt dualizeGammas()}, where for example the following
identity is used
$$
\g^{m_1...m_7}_{\alpha \beta} = - {1\over 3!}i \e_{m_1...m_7n_1...n_3}\g^{n_1...n_3}_{\a\b}.
$$
The cases where epsilon tensors are contracted with gamma matrices is also dealt with,
$$
\e^{m_1{\ldots} m_3n_1{\ldots} n_7}(\chi^1 \g_{n_1{\ldots} n_7}\chi^2) = 
5040\, i (\chi^1 \g^{m_1 m_2 m_3 }\chi^2).
$$
For completeness, the gamma matrix conventions \mukho\ are such that\foot{The signs change when both spinor indices of
the matrix matrices change from Weyl to anti-Weyl.}
$$
\g^{m_1{\ldots} m_9}_{\a\b} = i\,\e^{m_1{\ldots} m_9 n_1}(\g_{n_1})_{\a\b}, \quad
(\g^{m_1{\ldots} m_8})_{\a}^{{\hskip.04in}\b} = {1\over 2!}i\,\e^{m_1{\ldots} m_8 n_1{\ldots} n_2}(\g_{n_1{\ldots} n_2})_{\a}^{\hskip.04in\b}
$$
$$
\g^{m_1...m_7}_{\alpha \beta} = - {1\over 3!}i \e^{m_1...m_7n_1...n_3}\g^{n_1...n_3}_{\a\b},\quad
(\g^{m_1{\ldots} m_6})_{\a}^{{\hskip.04in}\b} = {1\over 4!}i\,\e^{m_1{\ldots} m_6 n_1{\ldots} n_4}(\g_{n_1{\ldots} n_4})_{\a}^{\hskip.04in\b},
$$
\eqn\gammadual{
\g^{m_1{\ldots} m_5}_{\a\b} = {1\over 5!}i\,\e^{m_1{\ldots} m_5 n_1{\ldots} n_5}(\g_{n_1{\ldots} n_5})_{\a\b}, \quad
\g^{m_1{\ldots} m_3}_{\a\b} = - {1\over 7!}i\,\e^{m_1{\ldots} m_3 n_1{\ldots} n_7}(\g_{n_1{\ldots} n_7})_{\a\b}
}
It is important to notice that {\tt FORM} uses the convention that
$
\e_{m_1{\ldots} m_{10}}\e^{m_1{\ldots} m_{10}} = 10!
$
instead of $-10!$, so that is why there are factors of $i$ together with epsilon tensors in {\tt PSS}.

The procedure {\tt fieldStrength()} substitutes $F^I_{mn} = k^I_m e^I_n - k^I_n e^I_m$ and {\tt onShell()}
annihilates terms with $(k^I\cdot k^I)$ and $(k^I\cdot e^I)$. If there are fermionic particles, the procedure
{\tt diracEquation()} uses the Dirac equation to reduce the rank of the gamma matrices when there is a momentum
of one of the particles in the fermionic bilinear being contracted with one of its indices, for example
$$
(\chi^1\g_{mnk_2}\chi^2) = k^2_m (\chi^1 \g_n \chi^2) - k^2_n (\chi^1\g_m \chi^2).
$$
Finally, the procedure {\tt momentumConservation()} applies the conservation of momentum to one of
the labels, which can be manually chosen by setting the ``dollar'' variable {\tt \$kn} in the beginning
of {\tt pss.frm} (if let at its default value of zero, then an automatic choice is made).

After the above (simplified) sequence of steps the desired component expansion of the 
superspace expression is printed on the screen. 

If one chooses the particles 1 and 2 to be fermionic by using {\tt \#define field1 1} etc, rerunning the program
results in,
\begintt
[mafra@Pilar:pss] tform -q -w2 pss.frm 

   [4pts_one-loop] =
       + 1/11520*chi1*ga(k4,e3,e4)*chi2*u
       + 1/11520*chi1*ga(k4,e3,e4)*chi2*t
       - 1/11520*chi1*ga(k4)*chi2*e3.e4*u
       + 1/11520*chi1*ga(k4)*chi2*e3.e4*t
       + 1/5760*chi1*ga(e3)*chi2*k1.e4*u
       - 1/5760*chi1*ga(e3)*chi2*k2.e4*t
       + 1/11520*chi1*ga(e4)*chi2*k4.e3*u
       - 1/11520*chi1*ga(e4)*chi2*k4.e3*t
      ;

Momentum conservation: k3 eliminated
Gauge invariance: not tested 
 
  0.08 sec + 0.42 sec: 0.50 sec out of 0.32 sec
\endtt

\subsec{Debug options}

There are three pre-defined debug options which can be used to check intermediate steps in the computation, {\tt sfexpand},
{\tt psonly} and {\tt nofierz}. If {\tt sfexpand} is invoked (using the {\tt -d} flag; see {\tt FORM}'s manual), 
then only the superfield expansion in
terms of $\t$'s is printed,
\begintt
[mafra@Pilar:pss] tform -q -w2 -d sfexpand pss.frm

[4pts_one-loop] =
   - 1/256*la*ga(e1)*th*la*ga(k4)*ga(m2,n2)*th*la*ga(n)*ga(m3,n3)*th*
     th*ga(n,m4,n4)*th*F2(m2,n2)*F3(m3,n3)*F4(m4,n4)
   - 1/384*la*ga(e1)*th*la*ga(m)*ga(k2,N1_?)*th*th*ga(m2,n2,N1_?)*th*
     la*ga(n)*ga(m3,n3)*th*F2(m2,n2)*F3(m3,n3)*F4(m,n)
   + 1/256*la*ga(e1)*th*la*ga(m)*ga(m2,n2)*th*la*ga(k4)*ga(m3,n3)*th*
     th*ga(m,m4,n4)*th*F2(m2,n2)*F3(m3,n3)*F4(m4,n4)
   + 1/384*la*ga(e1)*th*la*ga(m)*ga(m2,n2)*th*la*ga(n)*ga(k1,N1_?)*th*
     th*ga(m3,n3,N1_?)*th*F2(m2,n2)*F3(m3,n3)*F4(m,n)
   + 1/384*la*ga(e1)*th*la*ga(m)*ga(m2,n2)*th*la*ga(n)*ga(k2,N1_?)*th*
     th*ga(m3,n3,N1_?)*th*F2(m2,n2)*F3(m3,n3)*F4(m,n)
   + 1/384*la*ga(e1)*th*la*ga(m)*ga(m2,n2)*th*la*ga(n)*ga(k4,N1_?)*th*
     th*ga(m3,n3,N1_?)*th*F2(m2,n2)*F3(m3,n3)*F4(m,n)
   - 1/512*la*ga(N1_?)*th*th*ga(m1,n1,N1_?)*th*la*ga(m)*ga(m2,n2)*th*
     la*ga(n)*ga(m3,n3)*th*F1(m1,n1)*F2(m2,n2)*F3(m3,n3)*F4(m,n)
   ;
\endtt
which can be useful to check when something goes wrong. The debug option {\tt psonly}
prints the expression after the correlators were identified, for example,
\begintt
[mafra@Pilar:pss] tform -q -w2 -d psonly pss.frm

[4pts_one-loop] =
  - 1/128*F1(N1_?,N2_?)*F2(N3_?,N4_?)*F3(N5_?,N6_?)*F4(N3_?,N5_?)*
    ps111(N7_?,N4_?,N6_?,N7_?,N1_?,N2_?)
  - 1/256*F1(N1_?,N2_?)*F2(N3_?,N4_?)*F3(N5_?,N6_?)*F4(N3_?,N7_?)*
    ps311(N5_?,N6_?,N7_?,N8_?,N4_?,N8_?,N1_?,N2_?)
  - 1/256*F1(N1_?,N2_?)*F2(N3_?,N4_?)*F3(N5_?,N6_?)*F4(N5_?,N7_?)*
    ps311(N3_?,N4_?,N7_?,N8_?,N6_?,N8_?,N1_?,N2_?)
  - 1/512*F1(N1_?,N2_?)*F2(N3_?,N4_?)*F3(N5_?,N6_?)*F4(N7_?,N8_?)*
    ps331(N3_?,N4_?,N7_?,N5_?,N6_?,N8_?,N9_?,N9_?,N1_?,N2_?)
    ...
\endtt
where the other terms are similar and were omitted. Furthermore, {\tt nofierz} prints 
the superspace expansions before any Fierz manipulation is done,
\begintt
[mafra@Pilar:pss] tform -q -w2 -d nofierz pss.frm 

[4pts_one-loop] =
 + 1/24*chi1*ga(N1_?)*th*chi2*ga(N2_?)*th*la*ga(k2,N2_?,N3_?)*th*
   la*ga(N1_?)*th*la*ga(N4_?)*th*F3(N4_?,N5_?)*F4(N3_?,N5_?)
 + 1/48*chi1*ga(N1_?)*th*chi2*ga(N2_?)*th*la*ga(k2,N2_?,N3_?)*th*
   la*ga(N4_?,N5_?,N6_?)*th*la*ga(N1_?)*th*F3(N5_?,N6_?)*F4(N3_?,N4_?)
 + 1/24*chi1*ga(N1_?)*th*chi2*ga(N2_?)*th*la*ga(N1_?)*th*la*ga(k2)*th*
   la*ga(N3_?)*th*F3(N3_?,N4_?)*F4(N2_?,N4_?)
   ...
\endtt

There are many possible extensions and optimizations which can be made to {\tt PSS}, as it is available to
download at {\tt http://www.aei.mpg.de/\~{}crmafra/pss.tar.gz} under the GPL license. 
In particular, dealing with
four-fermion expansions is still not completely automated (nor guaranteed to be correct). It would be interesting to
implement the fermionic methods described in \stahn\ for this purpose. Furthermore, it should be straightforward to write
procedures to translate the full ten-dimensional components to four dimensions using the spinor helicity
formalism, in order to compare with the results appearing in \stieMHV. The possibilities are many and it is hoped that
{\tt PSS} provides a framework for further work.

\vskip 15pt {\bf Acknowledgements:} 
I want to thank Joost Hoogeveend for convincing me that {\tt PSS} may be useful to other
people and encouraging me to release it. I also thank Jos Vermaseren for his suggestions on how
to deal with generalized Kronecker deltas.
I acknowledge support by the Deutsch-Israelische
Projektkooperation (DIP H52). 

\listrefs

\end